\documentclass[11pt]{article}
\usepackage{amssymb,amsthm,amsmath,float,subfigure}
\usepackage{appendix}
\usepackage{booktabs} 
\usepackage{color}

\usepackage[total={6.5in,8.75in}, top=1.2in, left=0.9in, includefoot]{geometry}
\usepackage{graphicx}
\newcommand{\Eq}[1]{(\ref{eq:#1})}

\newcommand{\Sec}[1]{Section \ref{sec:#1}}
\newcommand{\Fig}[1]{Fig.~\ref{fig:#1}}

\newcommand{\Ass}[1]{Assertion~\ref{ass:#1}}

\newcommand{\InsertFig}[4]
{\begin{figure}[h!t]
 \centerline{
 \includegraphics[width=#4]{#1}
 }
 \caption{{\footnotesize #2}
 \label{fig:#3}}
\end{figure}}

\newcommand{\InsertFigTwo}[5] {
\begin{figure}[h!t]
 \centerline{
 \includegraphics[width=#5]{#1}
 \hskip 0.1in
 \includegraphics[width=#5]{#2}
 }
 \caption{{\footnotesize #3}
 \label{fig:#4}}
\end{figure}}

\newcommand{\InsertFigFour}[7] {
\begin{figure}[h!t]
 \centerline{
\renewcommand{\arraystretch}{0.01}
 \begin{tabular}{cc}
 \includegraphics[width=#7]{#1}& \includegraphics[width=#7]{#2} \\
 \includegraphics[width=#7]{#3} & \includegraphics[width=#7]{#4}
 \end{tabular}
 }
 \caption{{\footnotesize #5}
 \label{fig:#6}}
\end{figure}}

\newcommand{\bR}{{\mathbb{ R}}}

\newcommand{\bT}{{\mathbb{ T}}}

\newcommand{\cL}{{\cal L}}
\newcommand{\cO}{{\cal O}}

\newcommand{\cT}{{\cal T}}

\newcommand{\eps}{\varepsilon}

\newtheorem{ass}{Assertion}

\newcommand{\beq}[1]{\begin{equation}\label{eq:#1}}
\newcommand{\eeq}{\end{equation}}

\newenvironment{se}[1]{\equation\label{eq:#1}\aligned}{\endaligned\endequation}
\newcommand{\bsplit}[1]{\begin{se}{#1}}
\newcommand{\esplit}{\end{se}}

\title{Barriers to Transport and Mixing in Volume-Preserving Maps with Nonzero Flux}
\author{Adam M.~Fox\footnote{afox33@math.gatech.edu}
 \ and\ 
 Rafael de la Llave\footnote{rll6@math.gatech.edu}
	\ \thanks
 {
 AMF and RdlL  were supported in part by NSF grant DMS-1162544.  
 }
 \\
 	School of Mathematics\\
 	Georgia Institute of Technology \\
	Atlanta, GA 30332-0160 \\
}
\date{\today}
\begin{document}
\maketitle

\begin{abstract}
 \noindent
\emph{AMS Numbers:} 65P20 , 37J05, 37J40, 70H09 
\vspace*{1ex}\\
\noindent
In this paper we identify the geometric structures that restrict transport and mixing in perturbations of integrable volume-preserving systems with nonzero net flux.  Unlike KAM tori, these objects cannot be continued to the tori present in the integrable system but are generated by resonance and have a contractible direction.  We introduce a remarkably simple algorithm to analyze the behavior of these maps and obtain quantitative properties of the tori.  In particular, we present assertions regarding the distribution of the escape times of the unbounded orbits, the abundance of tori, and the size of the resonant regions.
\end{abstract}




\section{Introduction}\label{sec:Intro}

When considering flows in pulsating channels there are two natural questions:

\begin{enumerate}

\item

Is all of the material flushed out?

\item

Is the material thoroughly mixed?

\end{enumerate}
The goal of this paper is to numerically study some geometric structures that prevent a positive answer to these questions.  

We discover that the main objects that prevent transport are invariant tori of codimension one which cannot be continued to the integrable system and have contractible directions.  These tori, which we will refer to as \emph{secondary}, are generated by resonances.  

In this paper we use an exceedingly simple method to study secondary tori and their impact on transport and mixing in area and volume-preserving maps.  We generate random initial conditions and determine if the resulting orbits remain bounded.  We observe that the orbits that remain bounded are in regions bounded by secondary tori.  These regions contain other secondary tori as well as chaotic regions.  

\subsection{Preliminaries}

We consider measure-preserving mappings of $\bT^d \times \bR$ and focus on the most important physical cases which have $d=1,2$.  A simple example is 

\beq{VPMap}
(x',z')=f_{\lambda}(x,z)=(x+\Omega(z'), z+\lambda)
\eeq
with angles $x\in \bT^d$ and action $z\in\bR$.  Note that $\lambda$ can be interpreted as a mean flux.   If $\lambda \neq 0$ all the trajectories are unbounded and escape, however the system remains completely unmixed.  On the other hand, if $\lambda=0$ the flow is stratified.  Every orbit lies on an invariant torus $\cT_{z}=\bT^d \times \{z_0\}$ with constant action.  All tori homotopic to $\cT_{z}$ are referred to as \emph{rotational}.  Note that rotational tori separate the phase space into two regions.

A system with richer dynamics is the model proposed in \cite{Dullin12}
\beq{VPMap2}
(x',z')=f_{\eps,\lambda}(x,z)=(z+\Omega(z'), z+\eps g(x) + \lambda)
\eeq
where $g(x)$ is an average-zero function.  For simplicity, we assume $\Omega$ and $g$ are analytic.  When $\lambda=0$ there are KAM results \cite{delaLlave01,Xia92,Blass,Cheng90a,Cheng90b,Yoccoz92,Vayda12} that show that when $|\eps| \ll 1$ the model \Eq{VPMap2} possesses a set of invariant rotational tori of positive measure on which the dynamics is conjugate to a rigid rotation.  Orbits cannot cross these tori, hence all orbits remain bounded.  Moreover, they also prevent complete mixing since they separate regions of phase space.  

When $\lambda \neq 0$ rotational tori cannot exist regardless of the size of $\eps$.  As is well known, the mean flux is equivalent to the volume of the region bounded between a rotational torus and its iterate.  By definition, if the torus is invariant then this area, and therefore the mean flux, is zero.  

Nevertheless, as we will see, for $ |\eps| \ll 1$ the perturbations generate some new $d$-dimensional invariant structures that are not present in the integrable case and are contractible to lower dimensional tori.  These structures persist when the system is perturbed in both $\eps$ and $\lambda$, as can be established using KAM theory.  Since they are codimension one they separate regions of space.  Any orbits trapped within the torus cannot escape and therefore cannot become unbounded.  

Secondary tori in twist maps, often referred to as \emph{islands}, have been extensively studied -- see, for example, \cite{Duarte08,Gorodetski12,Roberto04,Simo97} -- and have been shown to exist for arbitrarily large perturbations \cite{Duarte94,Aubry92}.  In higher dimensional maps the secondary tori appear as \emph{tubes} \cite{Feingold88b}. In analogy with the 
case of area-preserving maps, these tubes 
are bounded by KAM tori.   It is an open question as to whether these tubes also  exist when the perturbation is very strong. A forthcoming paper by the authors \cite{Fox14b} establishes the existence of these tubes in near-integrable volume-preserving maps.

We will explore the effect of these secondary tori by adding a flux term to two oft-studied systems.  The Standard Area-Preserving Map,
\bsplit{StdMap}
x'&=x+z' \\
z'&=z+\frac{\eps}{2\pi}\sin(2\pi x) + \lambda,
\esplit
is an example of \Eq{VPMap2} with $d=1$.  It was introduced by Chirikov \cite{Chirikov79} as a model for the behavior of a generic Hamiltonian system near resonance (see also \cite{Escande85}). The map \Eq{StdMap} has also been used to model many natural phenomena, see \cite{Meiss92} for a comprehensive overview. Although \Eq{StdMap} is generally referred to as \emph{The Standard Map} in the literature, we will call it the Standard Area-Preserving Map in this paper to avoid confusion with other analogues in higher dimensions. 

The Standard Volume Preserving Map, \cite{Dullin12}
\bsplit{StdVP}
x_1'&=x_1+\gamma+z' \\
x_2'&=x_2-\delta+\beta(z')^2 \\
z'&=z+\eps(\sin(2\pi x_1) + \sin(2\pi x_2) +\sin(2\pi (x_1-x_2))) + \lambda,
\esplit
is of the form of \Eq{VPMap2} with $d=2$.  Following \cite{Meiss12a,Fox13a,Fox14a} we use the parameters $\beta=2$ and $\gamma=\tfrac12(-1+\sqrt{5})$ and set $\delta=0$.  Maps of this form naturally arise in the study of incompressible fluid flows subject to periodic perturbations. The volume preservation is a direct consequence of the physical properties of 
preservation of mass and incompressibility - see, for example,  \cite{Cartwright96,Feingold87,Feingold88a}.  The addition of the flux term $\lambda$ allows us to model these fluids moving through a channel, such as water in a pipe or blood through an artery subject to periodic perturbations.

In \Sec{Methods} we outline numerical techniques to analyze the dynamics of \Eq{StdMap} and \Eq{StdVP}.  The main observation is that a non-zero flux destroys all rotational invariant tori.  However, this flux does not destroy the secondary tori which bound regions of phase space.

The behavior of the unbounded orbits, those not enclosed by secondary tori, is described in \Sec{Unbounded}.  We provide evidence for the assertion that the escape time of the orbits can be modeled as a Gamma random variable whenever secondary tori are present.

In \Sec{Bounded} we describe the bounded orbits and the secondary tori that enclose them.  In particular we study the abundance of the tori in parameter space and the Lyapunov exponents and global rotation vectors of the trapped orbits.  We will also present an assertion on the region in parameter space for which secondary tori may exist and provide numerical estimates for the size of the resonant regions.  

\section{Numerical Methods}\label{sec:Methods}

The numerical methods we employ to study these maps are exceptionally simple.  We select random initial conditions uniformly on $0 \leq x ,z \leq1$ and iterate the map $f_{\eps,\lambda}$ a specified number of times, $n_{max}$, to generate an orbit.  We classify an orbit as unbounded if the action $z$ grows larger than some predetermined threshold $z_{th}$, i.e. if $|z|>z_{th}$.  Otherwise, the orbit is said to be bounded.  Both constants $z_{th}$ and $n_{max}$ must be carefully chosen. 

If the orbit is bounded we compute the maximal Lyapunov exponent
\beq{Lyap}
\cL=\lim_{n\to\infty} \log_{10}\frac{1}{n}||Df^{n}(x_n,z_n)\vec{v}||,
\eeq
where $Df$ is the Jacobian matrix and $\vec{v}$ is some random initial vector, and the global rotation vector $\omega \in \bR^d$
\beq{Rot}
\omega(x_0,z_0)=\lim_{n\to\infty} \frac{F_{x}^{n}(x_0,z_0)-x_{0}}{n}
\eeq
where $F_{x}$ are the angle coordinates of the lift of $f$ to the universal cover. Note that this global rotation vector should not be confused with the internal rotation vector.  If an orbit is unbounded we record the escape times $T_e$, or the number of iterations needed for the action to grow larger than $z_{th}$. 

\vspace{5mm}

\noindent \textbf{The Algorithm for fixed $\eps$ and $\lambda$}
\begin{enumerate}

\item Fix $z_{th}$, $n_{max}$

\item Randomly select initial condition $(x_0,z_0) \in [0,1]^{d+1}$

\item Generate orbit $(x_n,z_n)=f_{\eps,\lambda}(x_{n-1},z_{n-1})$, $n=1,2,\cdots,n_{max}$

\begin{itemize}

\item At each iteration compute the Jacobian $Df^n$ and update limit \Eq{Lyap}

\item At each iteration update the lift $F_x^n$ and the limit \Eq{Rot}

\item If $|z_n| > z_{th}$ end iteration, declare orbit \emph{unbounded}.  Record escape time $T_{e}=n$.

\end{itemize}

\item Declare orbit \emph{bounded}.  Record maximal Lyapunov Exponent $\cL$ and rotation vector $\omega$.    

\item Repeat steps (2) - (4) as often as desired.

\end{enumerate}
We also note that this algorithm provides a method to compute the secondary tori in the zero net-flux case by taking the limit with $\lambda \to 0$.  

The bounded orbits of the Standard  Area-Preserving Map \Eq{StdMap} and Standard Volume Preserving Map \Eq{StdVP} are shown in \Fig{PhasePlot} for 500 initial conditions, $z_{th}=2$, $n_{max}=10000$, and several values of $\eps$ and $\lambda$.  Every bounded orbit that we observe is contained within a secondary torus.  There are of course other phenomena that might cause an orbit to remain bounded, for example if the orbit is periodic.  However, the confinement by surfaces is the only known method that provides a positive measure of bounded orbits.  

The secondary tori of the positive-flux Standard Area-Preserving Map \Eq{StdMap}, shown in \Fig{PhasePlot}(a)-(b), are indistinguishable from the typical islands of area-preserving twist maps.   These arise in nested families - tori of the same topology that are contained within each other. The families of these nested tori are not foliations and contain gaps in which more complicated behavior happens. In particular, in the gaps of these families of tori we have found orbits with positive 
Lyapunov exponents  and chaotic regions confined by the bounding tori.  

The secondary tori in the Standard Volume-Preserving Map \Eq{StdVP} are analogous to the islands of twist maps, however are rotational in one direction.  These tori appear as ``tubes'' as shown in \Fig{PhasePlot}(c)-(d).  These tubes are nested, much like the islands in the Standard Area-Preserving Map, and may be oriented in many different directions.  

\InsertFigFour{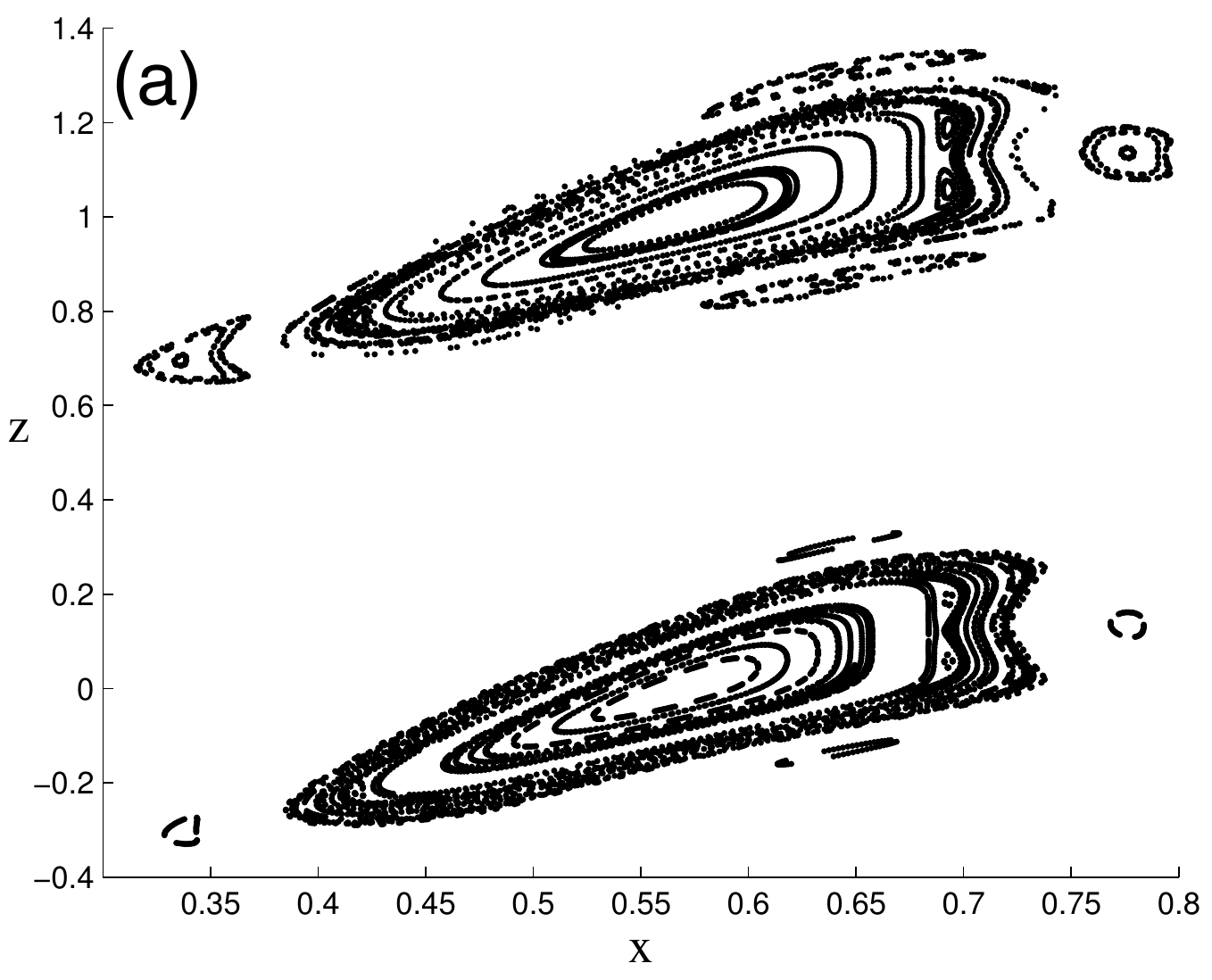}{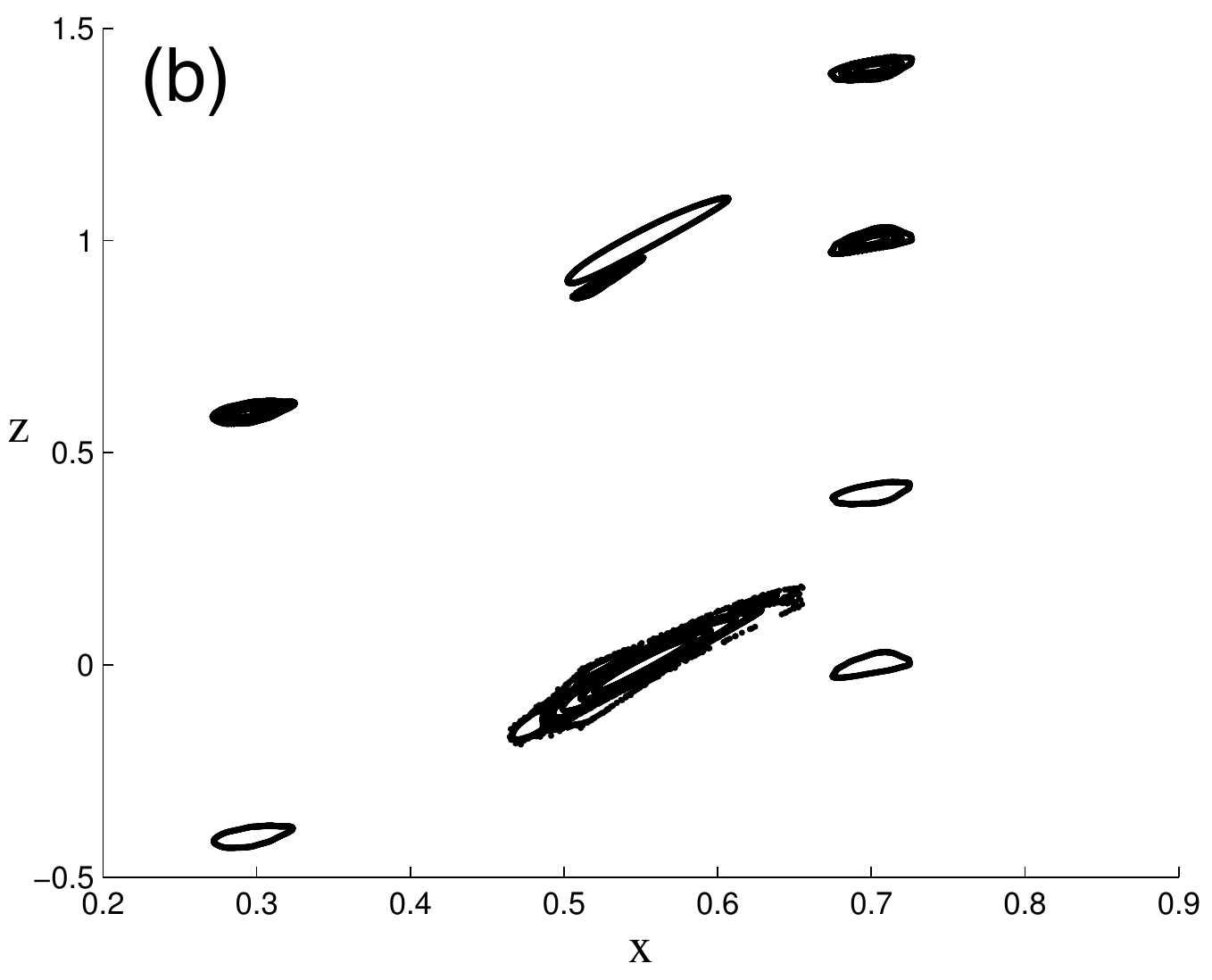}{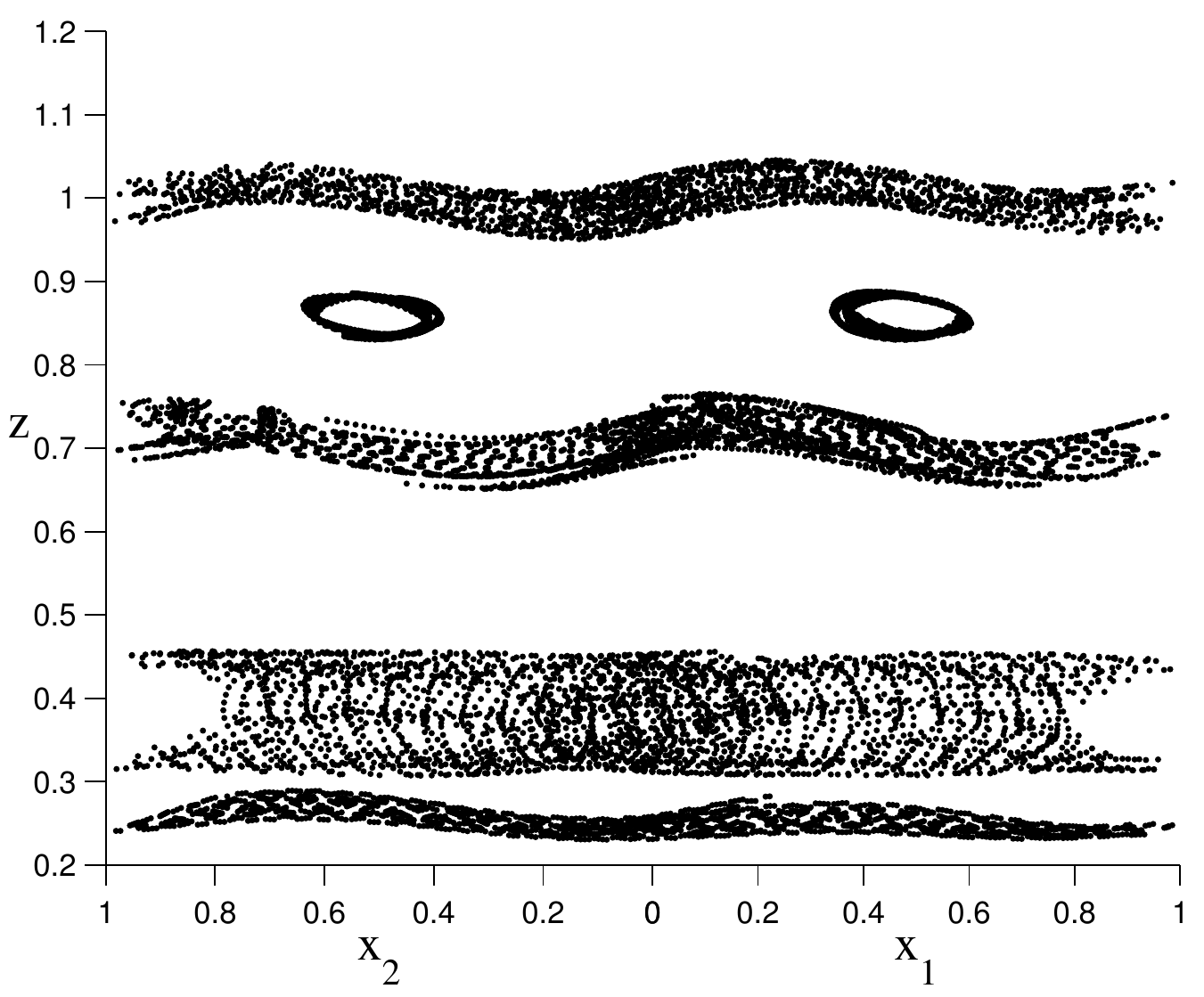}{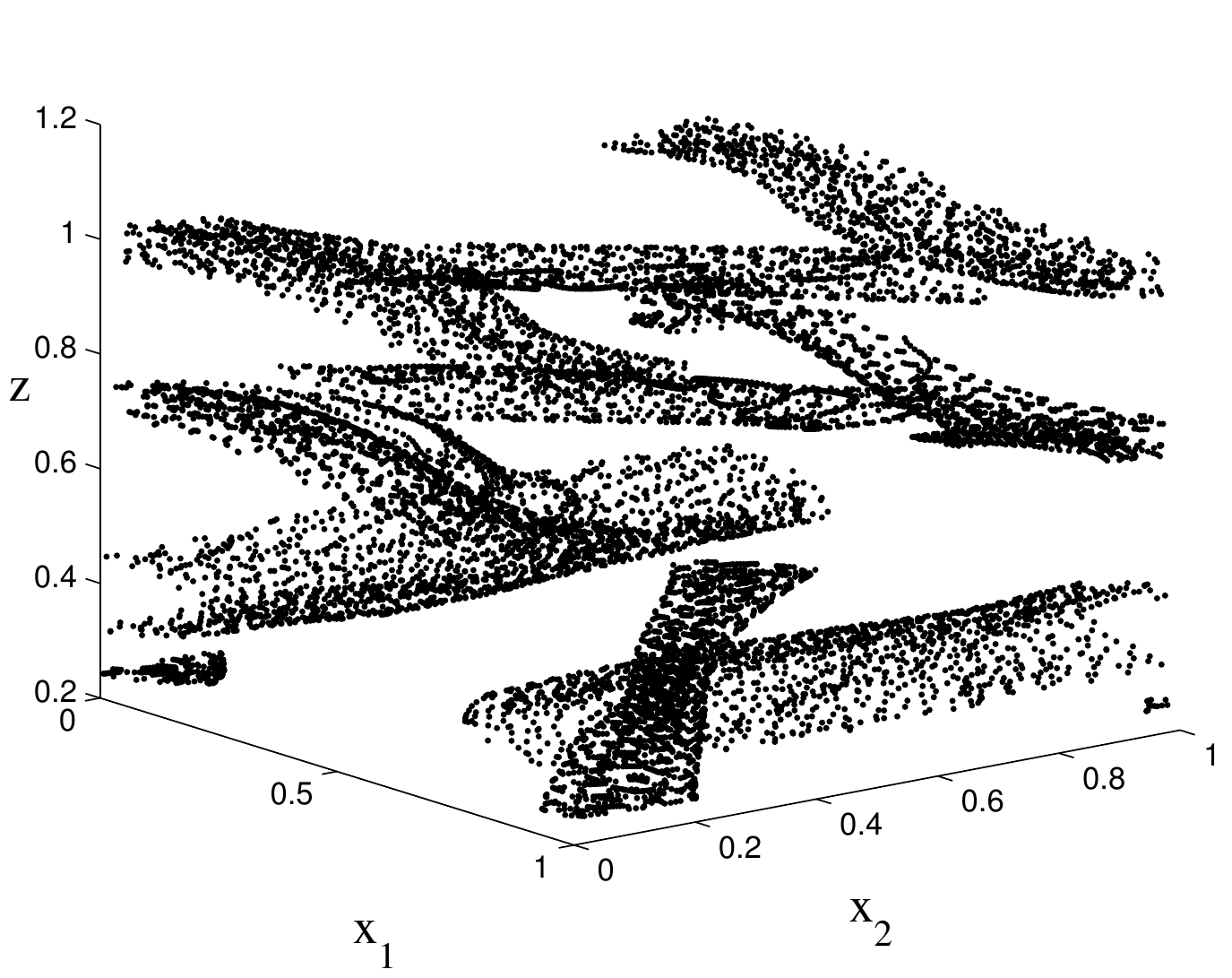}{Top Row: Bounded orbits of the Standard Area-Preserving  Map \Eq{StdMap} with $\lambda=0.2$ and Figure (a): $\eps=3$ and Figure (b): $\eps=4$. Bottom Row: Bounded orbits of the Standard Volume-Preserving Map \Eq{StdVP} for $\eps=0.016$ and $\lambda=0.001$.  All bounded orbits are contained within an invariant torus}{PhasePlot}{3in}

\section{Escape of Unbounded Orbits}\label{sec:Unbounded}

The algorithm described in \Sec{Intro} requires the user to specify the value of $z_{th}$ and $n_{max}$.  Before performing more complex experiments we must establish appropriate values for both parameters.  In this section we will examine the length of time needed for unbounded orbits to escape, i.e. for $|z|>z_{th}$.  This is a fundamentally important question because if the escape times tend to be very long then our algorithm will be time consuming and may potentially misclassify orbits.  If, however, the escape times are relatively short then we will be able to quickly and accurately perform our experiments.  

When $\eps=0$ the distribution of escape times is trivial.  The action $z$ evolves as $z_n=z_0+n\lambda$.  Since the initial points $z_0$ are uniformly distributed between 0 and 1 the escape times are uniformly distributed on $[\lambda^{-1}, \hspace{1mm} z_{th}\lambda^{-1}]$.  However, when $\eps \neq 0$ and secondary tori are present, the distribution of escape times fundamentally changes.  This new distribution has a heavy tail, which is not unexpected.  Indeed, this is consistent with measurements of transport in area-preserving twist maps \cite{Meiss92,Chirikov83,Karney83}.  If we ignore this heavy tail, we find that the escape times are Gamma distributed.

\begin{ass}\label{ass:Escape}
The tailless escape times $T_e$ of the unbounded orbits of maps of the form \Eq{VPMap2} are random variables with Gamma Distribution whenever secondary tori exist.  
\end{ass}

\noindent \emph{Evidence: }  \\
The Kolmogorov-Smirnov test \cite{Hogg10} may be used to determine whether the underlying distribution of a given sample of random variables is a reference distribution.   This is done by comparing the largest vertical distance between the empirical distribution function of the sample data
\[
\mbox{EDF}(x)=\frac{1}{n}\sum_{i=1}^n I_{X_{i}\leq x}
\]
(where $I_{X_{i}\leq x}=1$ if $X_i\leq x$ and 0 otherwise) and the cumulative distribution function of the reference distribution.  This distance is known as the Kolmogorov-Smirnov statistic can be used to compute the $p$-value for the test.  

To perform this test we first generate 500 escape times with $n_{max}=25000$.  Although some orbits require more time to escape  these are rare and, as we show below, do not have a substantial effect on the results.  We then disregard the largest escape times, as discussed further below, to compensate for the heavy tail of this distribution.  Assuming the tailless data is Gamma distributed, the maximum-likelihood estimates are given by 

\bsplit{}
\hat{\alpha}&=\frac{1}{2}(\log(\overline{T_e})-\overline{\log(T_e)})^{-1} \\
\hat{\beta}&=\overline{T_e} \hat{\alpha}^{-1},
\esplit
where $\overline{{\color{white} \cdot} \cdot {\color{white} \cdot }}$ indicates the sample mean.  

The best-fit Gamma distribution for a given sample of escape times was found by finding the parameters $\alpha,\beta$ that maximized the $p$ value of the Kolmogorov-Smirnov test.  This maximization was performed using the Nelder-Mead algorithm \cite{Nelder65}, seeded with the initial guess of $\hat{\alpha},\hat{\beta}$.  We employ the \emph{fminsearch} and \emph{kstest} MatLab commands to perform the search and compute the appropriate $p$-values.  

\subsection{Escape Times in the Standard Volume-Preserving Map}
The tail of the escape time data for the Standard Volume-Preserving Map \Eq{StdVP} was removed by disregarding the largest 5\% of the values.  The $p$-values of the Kolmogorov-Smirnov test for the resulting sample is shown in \Fig{StdVPPvals}.  The choice of escape threshold $z_{th}$ played an important role in these tests.  As $z_{th} \to \infty$ the $p$ values grew closer to one, indicating a better fit.  In \Fig{StdVPPvals} a clear improvement can be seen from when $z_{th}=2$ in \Fig{StdVPPvals}(a) to when $z_{th}=5$ in \Fig{StdVPPvals}(b).  The difference is especially pronounced when $\lambda$ is large.  The empirical distribution function of the escape times for $\eps=0.05$, $z_{th}=5.0$, and several values of $\lambda$ is shown in \Fig{EscapeTimeFit} along with the best-fit Gamma cumulative distribution functions.

\InsertFig{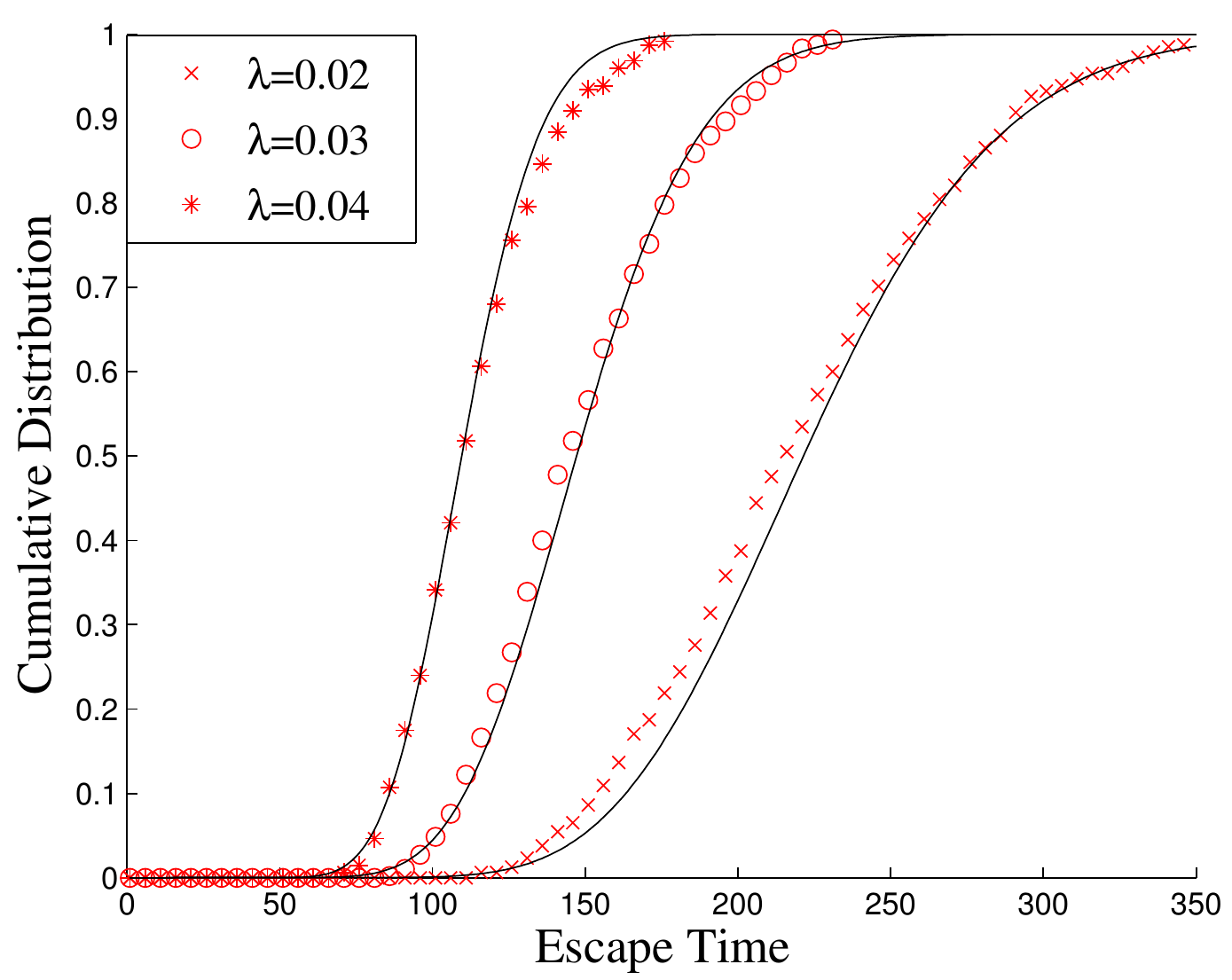}{Cumulative Distribution Function for the best-fit Gamma distribution (solid black line) and the tailless experimental data (red points) for the Standard Volume-Preserving Map with $\eps=0.06$.  The Gamma distributions have parameters $(\alpha,\beta)=(19.80,11.38)$ for $\lambda=0.2$, $(\alpha,\beta)=(22.20,6.73)$ for $\lambda=0.3$, and $(\alpha,\beta)=(30.4,3.65)$ for $\lambda=0.4$.  The corresponding $p$-values for the Kolmogorov-Smirnov test are 0.91, 0.73, and 0.67.}{EscapeTimeFit}{3in}

\InsertFigTwo{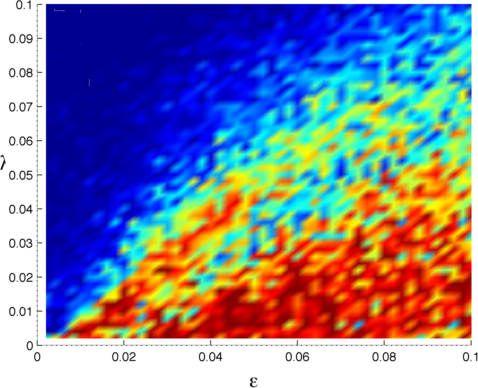}{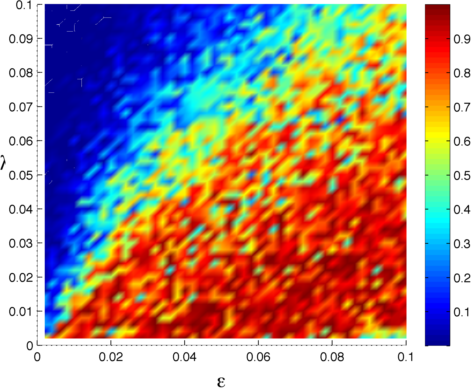}{$p$-values for the Kolmogorov-Smirnov test for the Standard Volume-Preserving Map \Eq{StdVP} with Figure (a): $z_{th}=2$ and Figure (b): $z_{th}=5$.}{StdVPPvals}{3in}

\subsection{Escape Times in the Standard Area-Preserving Map}

Removing the tail of the escape time data for the Standard Area-Preserving Map \Eq{StdMap} was significantly more complex.  This appears to be largely due to the different configurations of the tori in phase space.  For example, when the tori are wide and stretch across the $x$ dimension, such as in \Fig{PhasePlot}(a), the tail is relatively heavy.  However, when the tori are small, such as in \Fig{PhasePlot}(b), the tails tend to be shorter.  To compensate for these different behaviors three parameters were fit to maximize the $p$-value of the Kolmogorov-Smirnov test: the two parameters of the Gamma distribution and the percentage of largest escape times to disregard.  The $p$-values of the resulting tests with $z_{th}=50$ are shown in \Fig{StdMapPvals}(a) and the percentage of the points thrown out are shown in \Fig{StdMapPvals}(b).  The choice of $z_{th}$ once again proved important, with larger threshold values increasing the $p$-values especially for larger $\lambda$ values.  

\InsertFigTwo{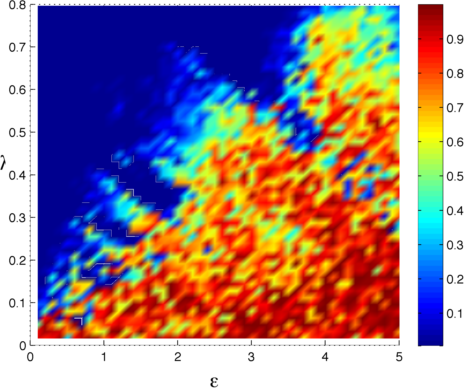}{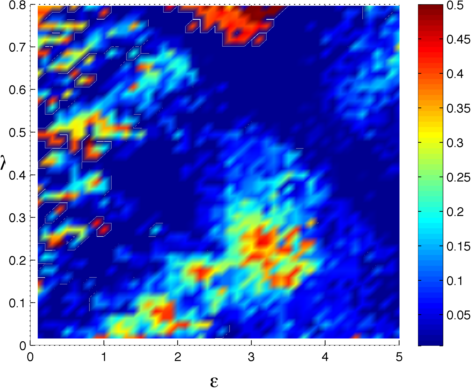}{Figure (a): $p$-values from the Kolmogorov-Smirnov test for the Standard Area-Preseving  Map \Eq{StdMap} with $z_{th}=50$ and fitting both the parameters of the Gamma distribution and the size of the tail to remove.  Figure (b): Percentage of the largest escape times discarded when performing the Kolmogorov-Smirnov test.}{StdMapPvals}{3in}

\subsection{Summary}
Whenever $\lambda \ll \eps$ or $\lambda \approx \eps$ the Kolmogorov-Smirnov test indicates that the tailless escape times are Gamma-distributed random variables.  As we show in \Sec{Bounded} this corresponds to the region for which secondary tori may exist.  Interestingly, even when $\lambda$ is slightly larger than this threshold we are able to fit the escape times to a Gamma distribution, however when $\lambda$ becomes significantly larger than $\eps$ the escape times are no longer Gamma distributed.  This might be explained by the existence of  remnant invariant sets that slow, but do not prevent, transport, analogous to the turnstiles that arise after invariant circles are destroyed \cite{Meiss86}.
\qed

\Ass{Escape} lends credibility to our numerical approach.  Since the escape times are Gamma distributed random variables we can easily compute the mean escape time, $\overline{T_e}=\alpha\beta$, where $\alpha$ and $\beta$ are the fitted parameters for the Gamma distribution.  For the Standard Area-Preserving Map with $z_{th}=50$ the largest mean escape time over all values of $\eps$ and $\lambda$ was computed to be 3107.4.  Similarly, the largest mean escape time for the Standard Volume-Preserving map over all values of $\eps$ and $\lambda$ was 2222 when $z_{th}=5$.  Although the escape times are not Gamma distributed when the secondary tori are absent these escape times are very rapid.

Since the unbounded orbits escape quickly the algorithm described in \Sec{Methods} will be able to distinguish between bounded and unbounded orbits with small $n_{max}$.  For the remainder of this paper we will use $z_{th}=2$ and $n_{max}=50000$ for the Standard Area-Preserving Map and $n_{max}=500000$ for the Standard Volume-Preserving Map.  Although it is possible that we may misclassify some orbits that have an atypically long escape time, these misclassifications will be rare.

\section{Existence of Secondary Tori}\label{sec:Bounded}
Recall from \Sec{Methods} that the only observed bounded orbits are those that are contained within a secondary torus.  In this section we will further study these bounded orbits and the secondary tori that contain them.  

We begin by exploring the relative measure of the set of bounded orbits by constructing a grid in $(\eps,\lambda)$ space.  At each pair of parameter values 5000 initial points were iterated in the Standard Area-Preserving Map \Eq{StdMap} while 15000 orbits were generated in the Standard Volume-Preserving Map \Eq{StdVP}.  The relative measure of the set of bounded orbits is shown below for both systems.  We observe that tori are most frequent for small values of $\eps$ and $\lambda$.  This is not surprising as large amounts of flux or force are expected to lead to the destruction of tori.  Perhaps the most prominent feature of \Fig{TorusFreq} is the absence of bounded orbits whenever the flux is significantly larger than the amplitude of the forcing.

\InsertFigTwo{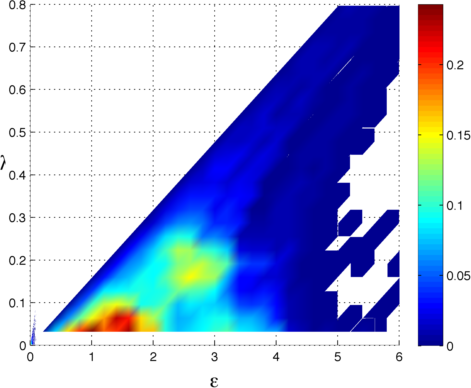}{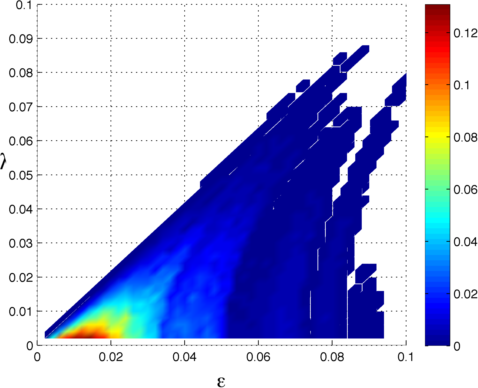}{Relative measure of the set of bounded orbits in Figure (a): the Standard Area-Preserving  Map \Eq{StdMap} and Figure (b): the Standard Volume-Preserving Map \Eq{StdVP}.  }{TorusFreq}{3in}

\begin{ass}\label{ass:EpsFlux}
Secondary tori only exist when
\begin{align}
\lambda &\leq \frac{\eps}{2\pi} \mbox{ in the Standard Area-Preserving Map \Eq{StdMap}}\nonumber \\
\lambda &\leq \eps+\eps^2\mbox{ in the Standard Volume-Preserving Map \Eq{StdVP}} \nonumber
\end{align}
\end{ass}

\noindent \emph{Evidence:}\\
To test this assertion the bisection method was used to establish to find $\lambda_{max}$, the smallest $\lambda$ for which no bounded orbits exist at fixed $\eps$.  The same number of points and iterations were used in the bisection methods as were used to generate \Fig{TorusFreq}.  For the Standard Area-Preserving Map \Eq{StdMap} we found $2\pi\lambda_{max}=\eps$ up to the accuracy of the bisection method.  Physically, this implies that the amplitude of the forcing must be at least as large as the amount of flux in order for secondary tori to exist.

Secondary tori can exist in the Standard Volume-Preserving Map \Eq{StdVP} when the flux is greater than the amplitude of the forcing.  The value of $\lambda_{max}$ is shown as a function of $\eps$ in \Fig{VPLastFlux}(a) along with dashed blue $\eps=\lambda$ line.  The difference between these values, $\lambda_{max}-\eps$, is plotted against $\eps$ on a logarithmic scale in \Fig{VPLastFlux}(b) where we can clearly a linear relationship.  Performing a least-squares fit yields the relation
\[
\lambda_{max}-\eps \approx 0.9757\eps^{1.9645} \approx \eps^2
\]
\qed

\InsertFigTwo{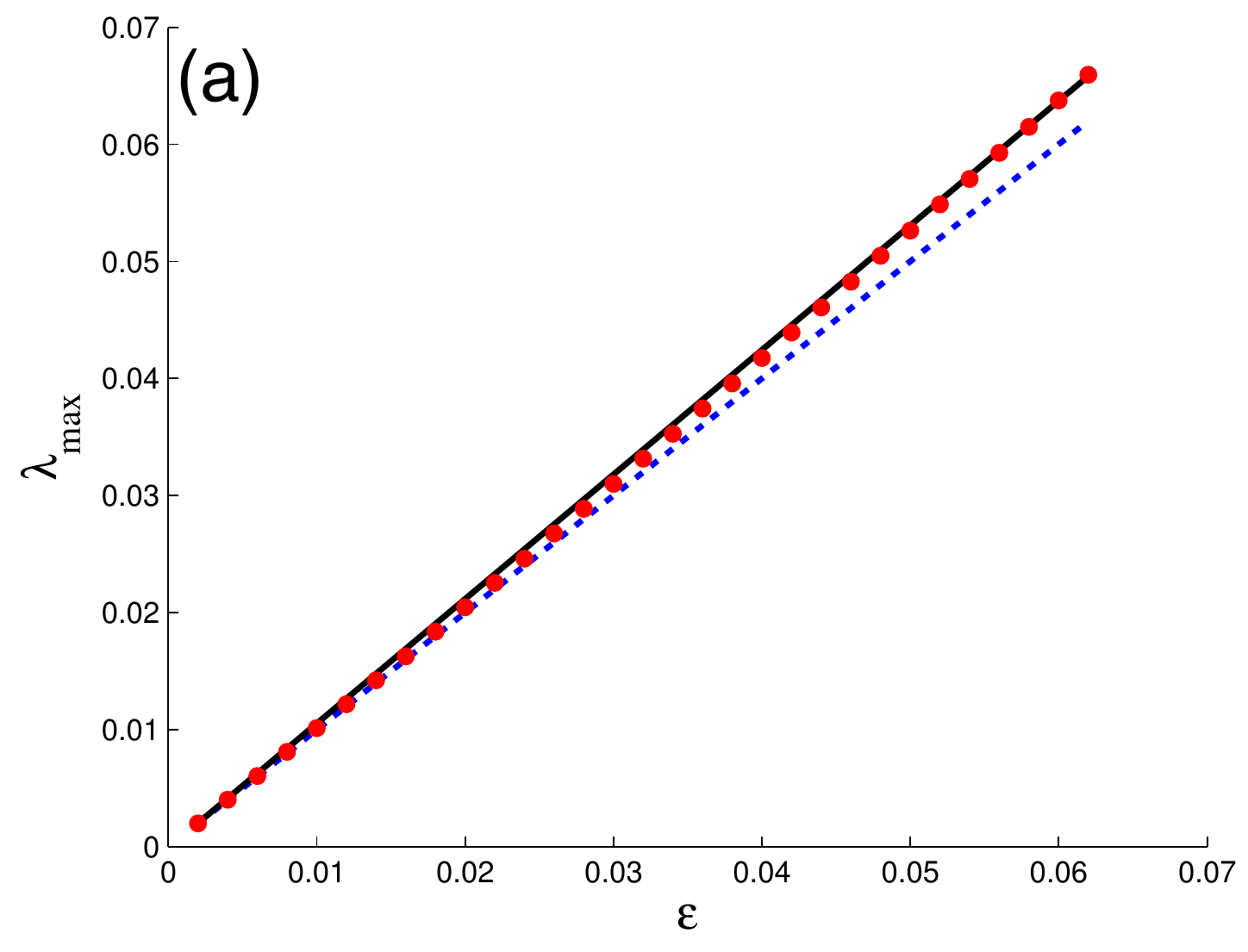}{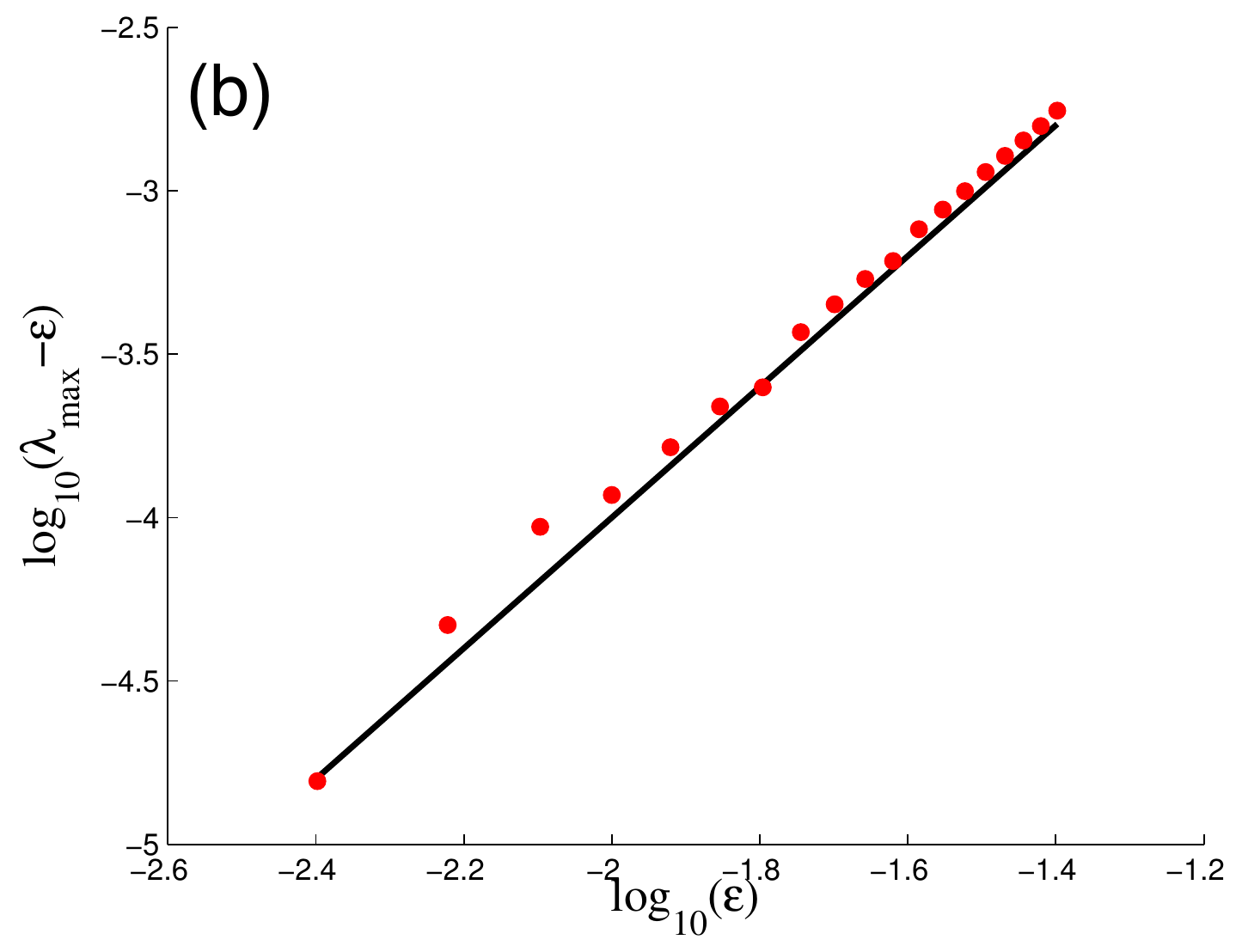}{Figure(a): The value of $\lambda$ such that no bounded orbits exist for fixed $\eps$ in the Standard Volume-Preserving Map \Eq{StdVP}.  The blue dashed line indicates where $\lambda=\eps$ while the solid black line shows $\lambda=\eps+\eps^2$.  Figure(b): The difference $\lambda_{max}-\eps$ as a function of $\eps$ plotted on a log-log scale.  A least-squares fit suggests that this difference grows as $\approx \eps^2$}{VPLastFlux}{3in}

We note that the estimate $\lambda_{max}=\eps+\eps^2$ is only a numerical estimate based on our observations.  There are likely higher order terms to this bound.  Secondly, all secondary tori may be destroyed prior to this value of $\lambda$, especially for larger values of $\eps$.  For other maps of the form \Eq{VPMap2} we expect
\[
\lambda_{max}=A_1\eps+A_2\eps^2 + \cdots
\]
where the constants $A_i$ depend on the structure of the map and force $g(x)$. 

The secondary tori in maps of the form \Eq{VPMap2} generally arise in nested families such as those seen in \Fig{PhasePlot}. The decline in the relative measure of the set of bounded orbits can be caused by either the destruction of the outermost layer or a decrease in its size.  Similarly, an increase in the relative measure can be caused by a growth in the outermost torus or the creation of a new torus that encompasses the existing tori.  \Fig{FreqSliceSmall} and \Fig{FreqSliceBig} shows this change for fixed $\eps$ and $\lambda$ in the Standard Area-Preserving Map \Eq{StdMap} using 50000 iterations and 500000 initial points.   

When $\eps$ is small, shown in \Fig{FreqSliceSmall}(a), bounded orbits are bountiful for small $\lambda$.  As $\lambda$ grows the number of bounded orbits decreases gradually.  We observe that this corresponds to the outermost layer of tori being destroyed. Note that this change is not monotonic - there are several instances where the outermost layer of tori grows or reforms, corresponding to an increase in the frequency of bounded orbits.

When $\lambda$ is small, shown in \Fig{FreqSliceSmall}(b), there are no bounded orbits for small $\eps$, recall \Ass{EpsFlux}.  However, once $\eps > 2\pi\lambda$, the number of bounded orbits quickly grows.  This is due to a growth in the size of the resonant region, as discussed in detail in \Sec{ResSize}.  As $\eps$ continues to grow a change occurs and the measure of the set of bounded orbits declines.  Once again, we observe a gradual decrease corresponding to the outermost layer of tori being peeled off.

\InsertFigTwo{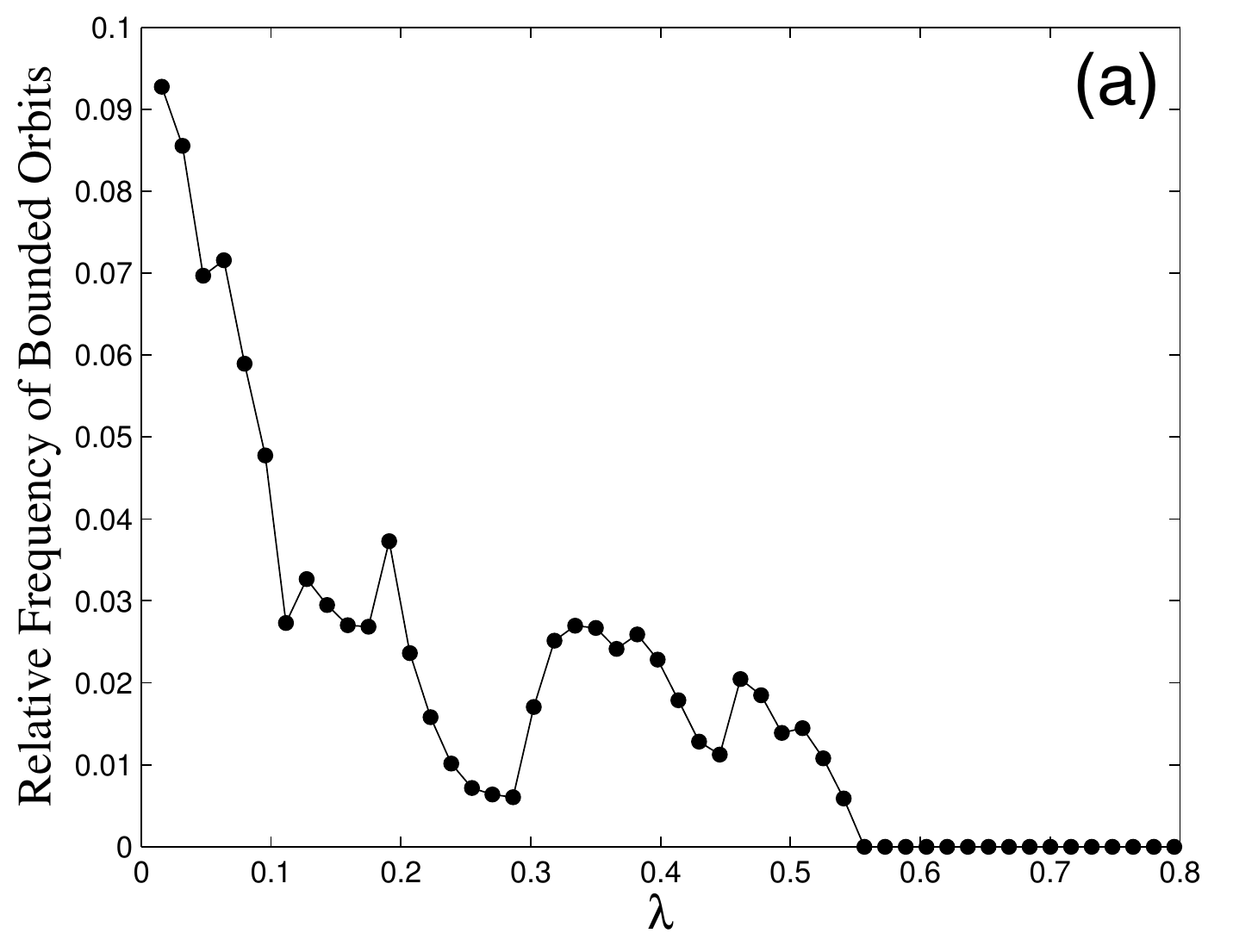}{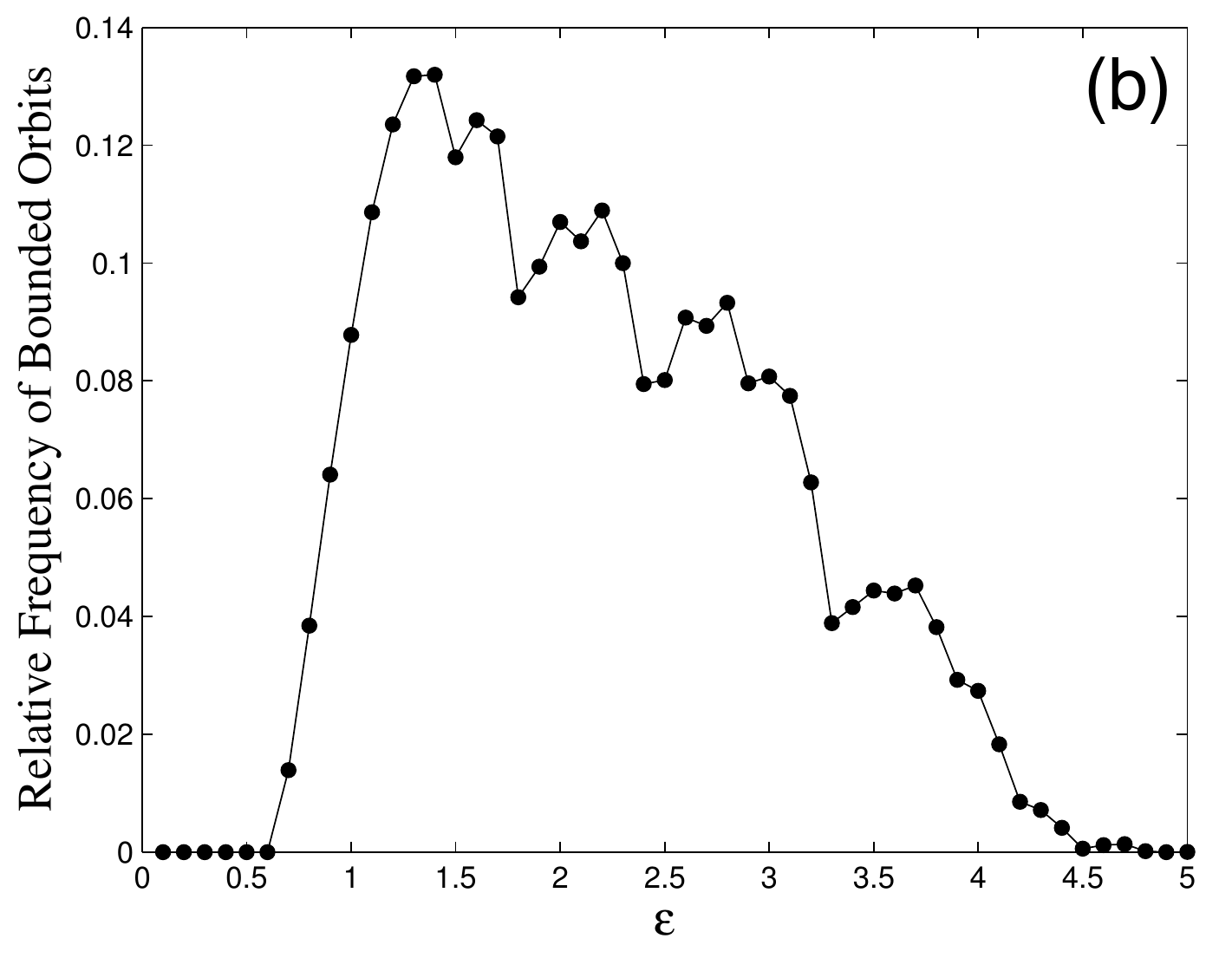}{Relative measure of the set of bounded orbits in the Standard Area-Preserving Map \Eq{StdMap} for Figure (a): $\eps=3.5$ as a function of $\lambda$ and Figure (b): $\lambda=0.1$ as a function of $\eps$.}{FreqSliceSmall}{3in}

When $\eps$ is large, shown in \Fig{FreqSliceBig}(a), the there are many tori for $\lambda \ll 1$, however these tori are quickly destroyed as $\lambda$ grows.  When $2\pi\lambda \approx \eps$ a second family of tori emerges, disappears, then reemerges as $\lambda$ grows.  When $\lambda$ is large there are no bounded orbits until $\eps>2\pi\lambda$ at which point the tori grow quickly.  These tori go through rapid growth and contraction, similar to the observed behavior for large $\eps$.  

\InsertFigTwo{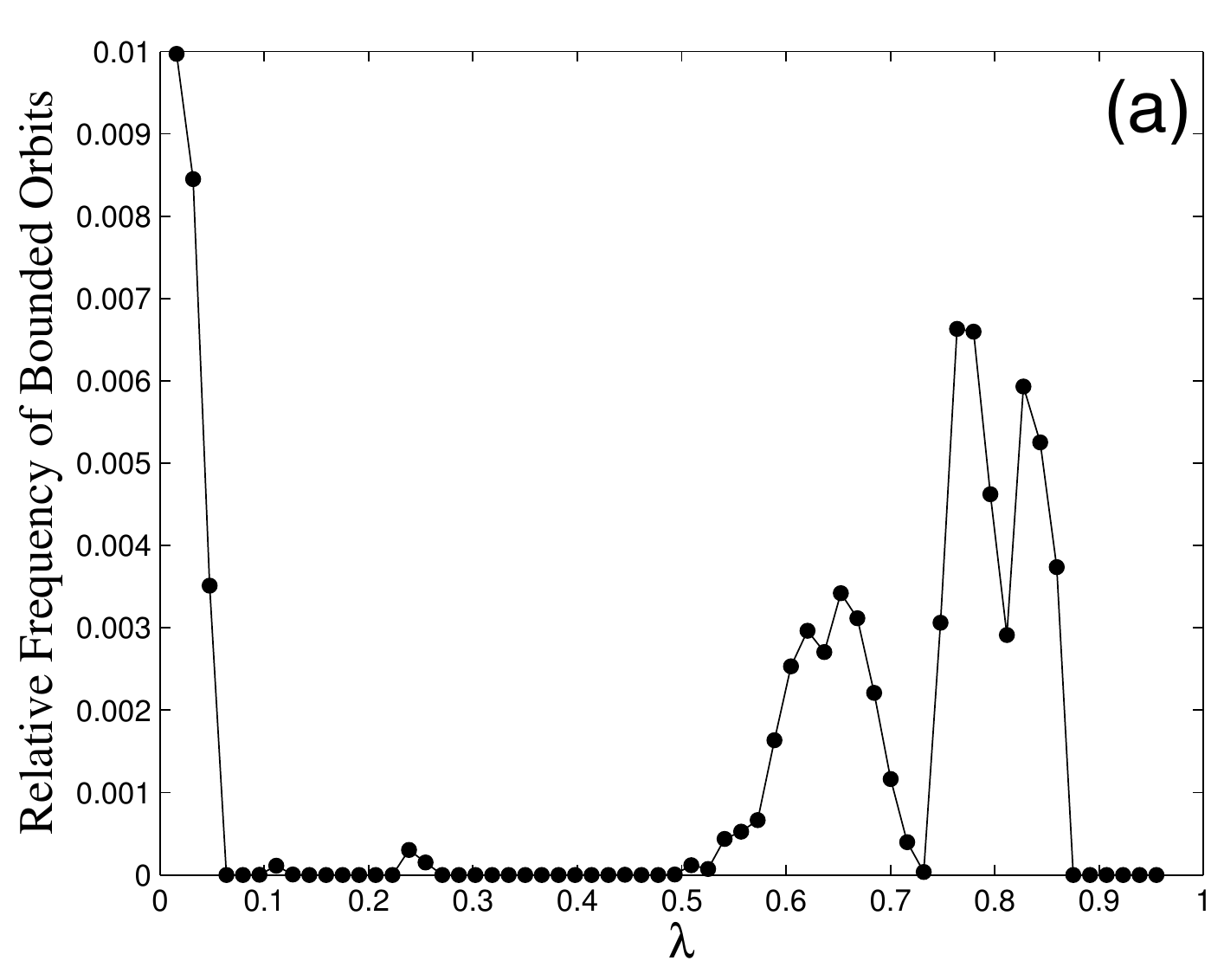}{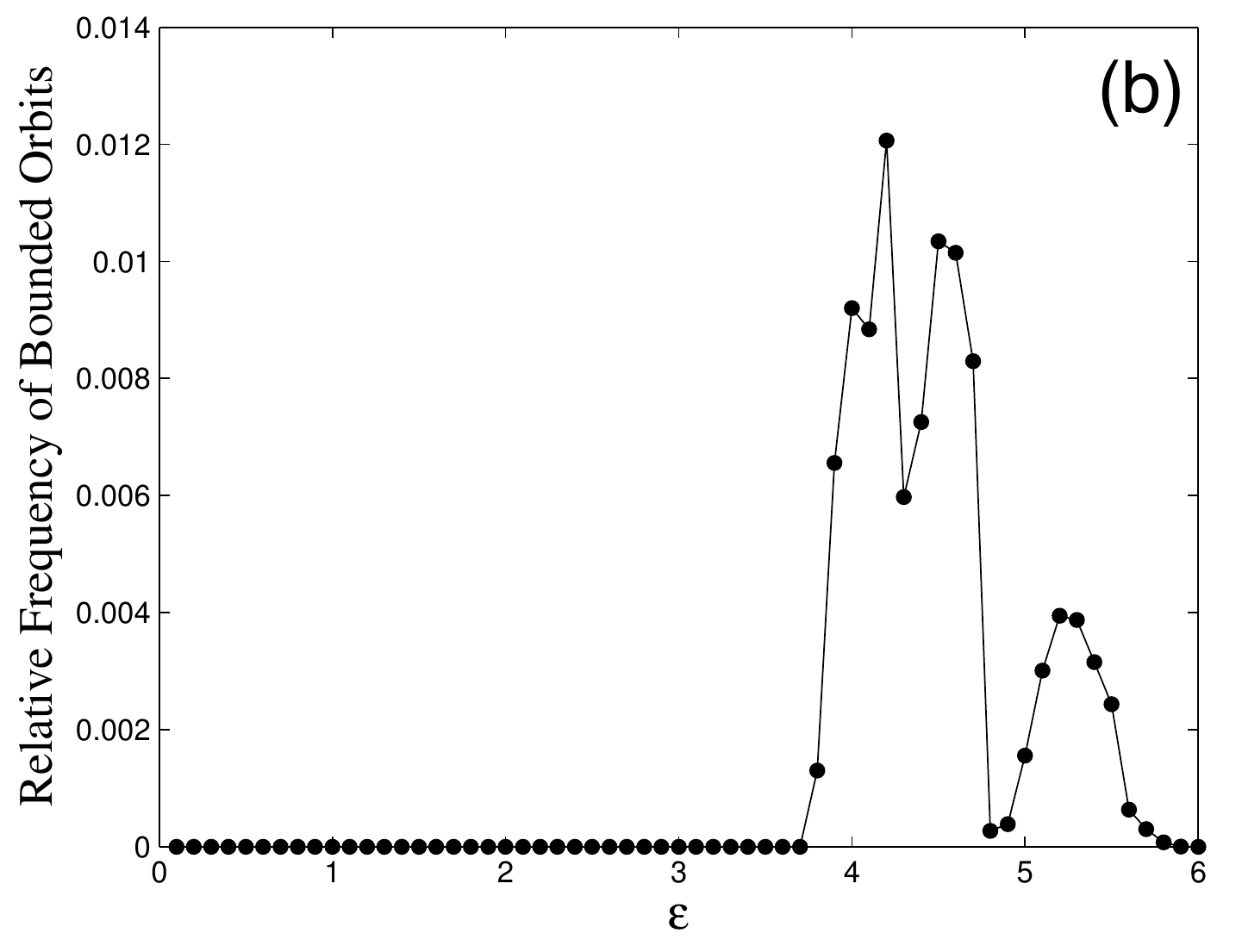}{Relative measure of the set of bounded orbits in the Standard Area-Preserving Map \Eq{StdMap} for Figure (a): $\eps=5.5$ as a function of $\lambda$ and Figure (b): $\lambda=0.6$ as a function of $\eps$. }{FreqSliceBig}{3in}

\subsection{Lyapunov Exponents of Bounded Orbits}\label{sec:Lyap}

The maximal Lyapunov exponents of the bounded orbits were estimated using \Eq{Lyap} on the same grid of $(\eps,\lambda)$ values as in \Fig{TorusFreq} using the same number of initial conditions and iterations.  The largest of the maximal exponents at each parameter value is shown in \Fig{LargeLyap}.  

\InsertFigTwo{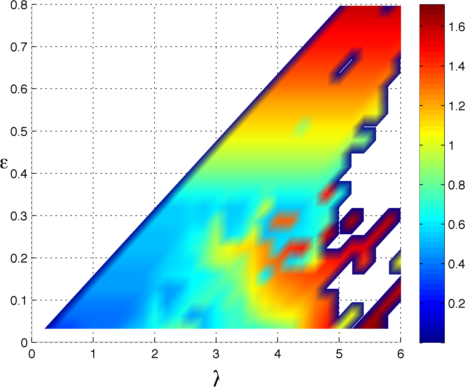}{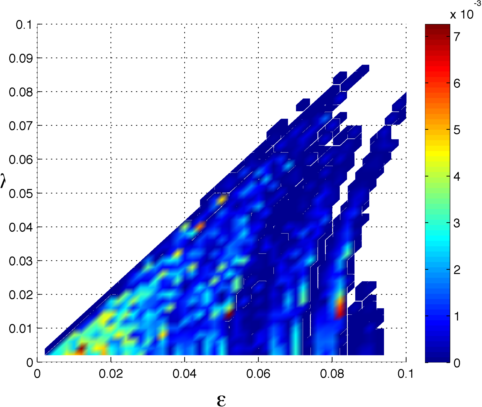}{Largest Lyapunov Exponents of the bounded orbits of the Figure(a): Standard Area-Presering  Map \Eq{StdMap} and Figure (b): Standard Volume-Preserving Map \Eq{StdVP}.  The largest exponents appear to occur near regions in parameter space for which no tori exist.}{LargeLyap}{3in}

\subsection{Global rotation vectors of orbits trapped by secondary tori}\label{sec:RotNum}

When $\eps=0$ all orbits of the Standard Volume-Preserving Map \Eq{StdVP} must have rotation numbers that lie on the parabola defined by the frequency map 
\[
\Omega(z)=(z+\gamma,-\delta+\beta z^2).
\]  
We approximate the rotation vectors of the bounded orbits with \Eq{Rot}.  These approximations are shown along with the frequency map $\Omega$ in \Fig{RotNumsVP}.  We observe that all of the secondary tori have global rotation vectors along the frequency map implying that these tori continue from the original resonance rather than being generated by higher order terms.  Similarly, all bounded orbits in the Standard Area-Preserving Map \Eq{StdMap} have rotation numbers near the low-order resonances $0$, $\tfrac12$, or $1$.   

\InsertFig{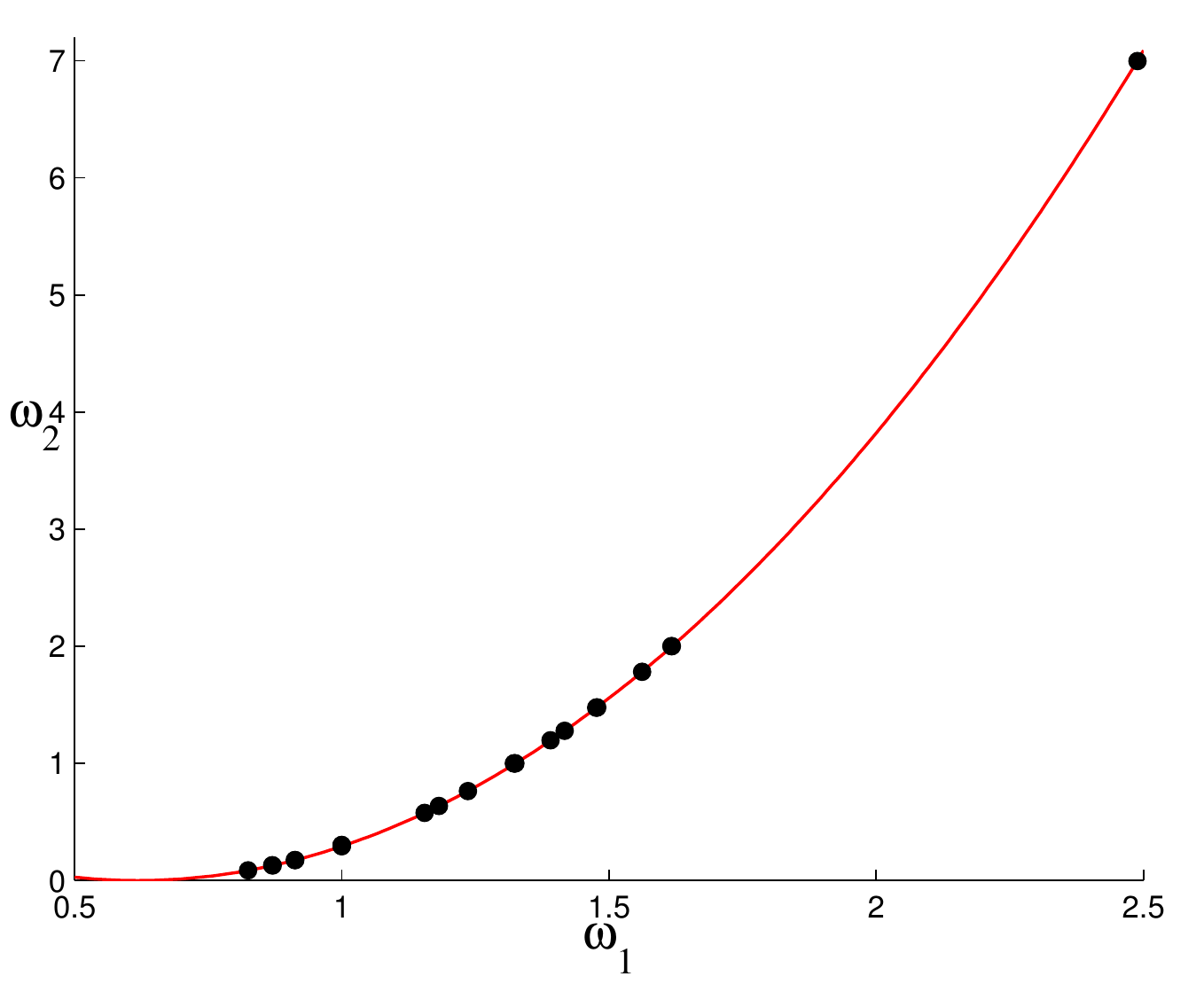}{Frequency map $\Omega(z)$ (red line) with computed frequencies of secondary tori (black dots)}{RotNumsVP}{3in}

\subsection{Size of the Resonant Region}\label{sec:ResSize}

The size of the resonant region as a function of $\eps$ can be approximated by measuring the diameter of the set bounded orbits, $D(\eps)$.  For the Standard Area-Preserving Map \Eq{StdMap}, we compute the Euclidean distance between every pair of points on the bounded orbits in a specified resonant region.  The largest distance over all bounded orbits at fixed $\eps$ is said to be the diameter of the resonance $D(\eps)$.  A similar process is used for the Standard Volume-Preserving Map \Eq{StdVP}, however we take a thin slice of a given of tube and only consider the distance between points in this slice.


\begin{ass}\label{ass:ResSize}
The size of the resonant regions $D(\eps)$ in maps of the form \Eq{VPMap2} is approximately
\beq{ResFit}
D(\eps)= A (\eps - B \lambda)^{1/2}
\eeq
for some nonnegative constants $A$ and $B$ and $\eps - B \lambda \ll 1$.  
\end{ass}


\noindent\emph{Evidence: } \\
The resonant region with $\omega \approx 0$ in the Standard Area-Preserving Map \Eq{StdMap} with $\lambda=0.25$ was measured for $\eps-2\pi\lambda\ll1$.  Recall from \Ass{EpsFlux} that this region corresponds to the smallest $\eps$ values for which secondary tori will exist.  At each value of $\eps$ 15000 initial conditions were chosen such that $|z| \leq 0.05$ and iterated 250000 times.  If the orbit remained bounded the first 200 points were employed to compute $D(\eps)$. 

A similar approach was used in the Standard Volume-Preserving Map \Eq{StdVP} with $\lambda=0.005$.  In this case the resonant region with $\omega \approx (\phi,2)$ where $\phi=\tfrac12(1+\sqrt{5})$ is the golden mean was explored.  The same number of orbits and iterations were used as in the Standard Area-Preserving Map case, however all initial conditions satisfied $x_2\in[0,0.5]$, $z\in [0.9,1.1]$.  These orbits were confined in a secondary torus that stretched across the $x_1$ dimension.  The diameter $D(\eps)$ was therefore computed within the slice $0\leq x_1 \leq 0.01$.   

The diameter of the resonances are shown in \Fig{ResSize} as a function of $\eps$.  The growth of the resonance can be modeled by \Eq{ResFit}.  Appropriate values must be computed for both $A$ and $B$.  \Ass{EpsFlux} implies that $B ={2\pi}$ for the Standard Area-Preserving  Map \Eq{StdMap} and $B=\eps+\eps^2 \approx \eps$ for the Standard Volume-Preserving Map.  The values of $A$ and $B$ were estimated from the experimental data using MatLab's nonlinear least-squares \emph{fit} command.  For the Standard Area-Preserving Map \Eq{StdMap} $B$ was found to equal $2\pi$ to within 8 digits while $A=0.5176$.  The data for the Standard Volume-Preserving Map \Eq{StdVP} gave the estimates $B=1$ to 12 digits and $A=26.89$.  Note that since $\lambda=0.005$ in this case \Ass{EpsFlux} implies $\lambda_{max}=\eps+\eps^2=\eps+\cO(10^{-6})$ hence it is not surprising that the fit gave $B=1$.
\qed

\InsertFig{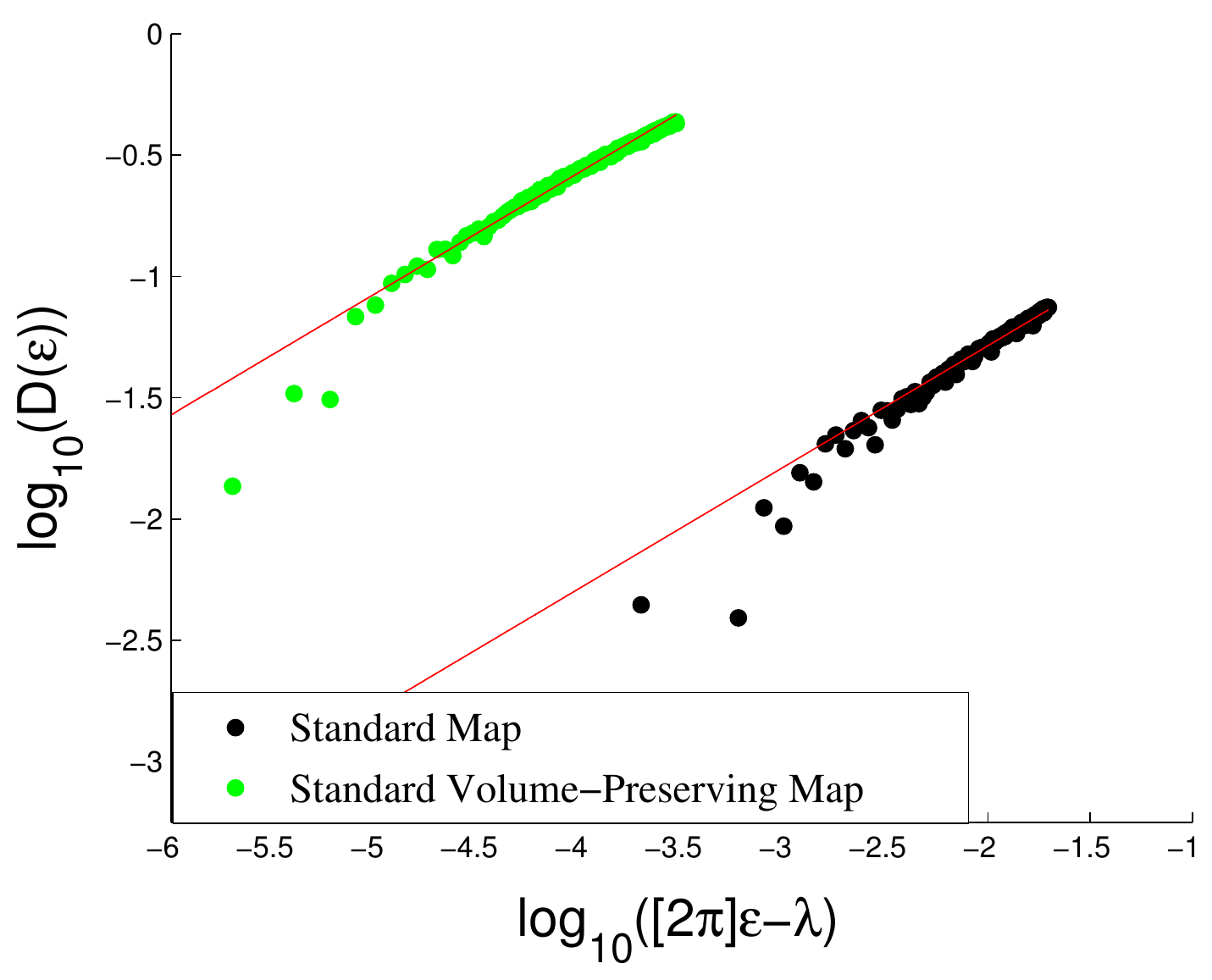}{Largest diameter of a bounded orbit, $D(\eps)$, plotted on a logarithmic scale.  The black data points in the lower right are for the Standard Area-Preserving Map \Eq{StdMap} with $\omega \approx 0$.  The best-fit curve, shown in red, was found to be $D(\eps)=0.5176(2\pi\eps-\lambda)$.  The green data points in the top right are for the Standard Volume-Preserving Map \Eq{StdVP} with $\omega \approx (\phi,2)$.  The best-fit curve for this case was $D(\eps)=26.89(\eps-\lambda)$}{ResSize}{4in}

\section{Conclusion}

In this paper we explored the dynamics of area and volume-preserving maps with positive flux.  Although rotational tori do not exist in these systems we have shown that secondary tori are present and play a fundamental role in the dynamics.  We provided evidence in support of assertions describing the size of the resonant regions and the parameter values for which secondary tori may exist.  We also demonstrated how these tori affect the rate of transport of the unbounded orbits.

There is significant potential for future research in this field.  
\begin{enumerate}

\item 

Can we construct a transport model that explains the Gamma distributed escape times described in \Ass{Escape}?

\item 

Can the claim that tori cannot exist in the Standard Area-Preserving Map \Eq{StdMap} if the net flux is greater than the amplitude of the forcing be rigorously proven?  

\item

Can the threshold $\lambda_{max}=\eps+\eps^2$ in the Standard Volume Preserving Map \Eq{StdVP} be improved and rigorously proven?

\item

Can a perturbative argument be used to prove \Ass{ResSize} and provide estimates for the parameters $A$ and $B$ in \Eq{ResFit}?

\item 

Can we numerically compute these secondary tori and predict their destruction under perturbation?

\item

In the two dimensional case, there has been a great deal of interest in finding ``the last invariant torus" in a confining island and they have been shown to have very interesting properties \cite{Simo97}.  It would be interesting to find similar results for the last tori in tubes.  

\item

Can these results be employed to study physical problems such as the build-up of cholesterol in arteries or the clogging of water pipes?

\end{enumerate}

\section{Acknowledgements}
The contributions of our undergraduate research assistants, Sally Blair and Ernesto Vargas, are gratefully acknowledged as are the many useful conversations with Prof.  James D. Meiss.

\bibliographystyle{alpha}
\bibliography{Channelflow}{}

\begin{thebibliography}{RdSCV04}

\bibitem[Aub92]{Aubry92}
Serge~J. Aubry.
\newblock The concept of anti-integrability: definition, theorems and
  applications to the standard map.
\newblock In {\em Twist mappings and their applications}, volume~44 of {\em IMA
  Vol. Math. Appl.}, pages 7--54. Springer, New York, 1992.
\newblock http://dx.doi.org/10.1007/978-1-4613-9257-6\_2.

\bibitem[BdlL]{Blass}
Timothy Blass and Rafael de~la Llave.
\newblock {KAM} theory for volume-preserving maps.
\newblock In Progress.

\bibitem[CFP96]{Cartwright96}
J.H.E. Cartwright, M.~Feingold, and O.~Piro.
\newblock Chaotic advection in three dimensional unsteady incompressible
  laminar flow.
\newblock {\em J. of Fluid Mech.}, 316:259--284, 1996.
\newblock http://dx.doi.org/10.1017/S0022112096000535.

\bibitem[Chi79]{Chirikov79}
B.V. Chirikov.
\newblock A universal instability of many-dimensional oscillator systems.
\newblock {\em Phys. Rep.}, 52:265--379, 1979.

\bibitem[Chi83]{Chirikov83}
B.V. Chirikov.
\newblock Chaotic dynamics in hamiltonian systems with divided phase space.
\newblock In L.~Garrido, editor, {\em Dynamical Systems and Chaos, Lecture
  Notes in Physics}, volume 179 of {\em Proc. Sympos. Pure Math.}, pages
  29--46. Springer-Verlag, Berlin, 1983.

\bibitem[CS90a]{Cheng90a}
C-Q. Cheng and Y.-S. Sun.
\newblock Existence of invariant tori in three-dimensional measure-preserving
  mappings.
\newblock {\em Celestial Mech. and Dyn. Astron.}, 47(3):275--292, 1990.
\newblock http://dx.doi.org/10.1007/BF00053456.

\bibitem[CS90b]{Cheng90b}
C.-Q. Cheng and Y.-S. Sun.
\newblock Existence of periodically invariant curves in 3-dimensional
  measure-preserving mappings.
\newblock {\em Celestial Mech. and Dyn. Astron.}, 47:293--303, 1990.
\newblock http://dx.doi.org/10.1007/BF00053457.

\bibitem[DLL01]{delaLlave01}
R.~De~La~Llave.
\newblock A tutorial on {KAM} theory.
\newblock In {\em Smooth ergodic theory and its applications (Seattle, WA,
  1999)}, volume~69 of {\em Proc. Sympos. Pure Math.}, pages 175--292. Amer.
  Math. Soc., Providence, 2001.

\bibitem[DM12]{Dullin12}
H.R. Dullin and J.D. Meiss.
\newblock Resonances and twist in volume-preserving mappings.
\newblock {\em Siam J. Dyn. Sys.}, 11:319--359, 2012.
\newblock http://dx.doi.org/10.1137/110846865.

\bibitem[Dua94]{Duarte94}
Pedro Duarte.
\newblock Plenty of elliptic islands for the standard family of area-preserving
  maps.
\newblock {\em Ann. Inst. H. Poincar\'e Anal. Non Lin\'eaire}, 11(4):359--409,
  1994.

\bibitem[Dua08]{Duarte08}
P.~Duarte.
\newblock Elliptic isles in families of area-preserving maps.
\newblock {\em Ergodic Theory and Dynamical Systems}, 28(06):1781--1813, 2008.

\bibitem[Esc85]{Escande85}
D.~F. Escande.
\newblock Stochasticity in classical {H}amiltonian systems: universal aspects.
\newblock {\em Phys. Rep.}, 121(3-4):165--261, 1985.

\bibitem[FdlL14]{Fox14b}
Adam~M. Fox and Rafael de~la Llave.
\newblock Deformation theory in volume-preserving maps and its application.
\newblock 2014.
\newblock In Preparation.

\bibitem[FKP87]{Feingold87}
M.~Feingold, L.P. Kadanoff, and O.~Piro.
\newblock A way to connect fluid dynamics to dynamical systems: Passive
  scalars.
\newblock In A.J. Hurd, D.A. Weitz, and B.B Mandelbrot, editors, {\em Fractal
  Aspects of Materials: Disordered Systems}, pages 203--205. Materials Research
  Society, Pittsburgh, 1987.

\bibitem[FKP88a]{Feingold88a}
M.~Feingold, L.P. Kadanoff, and O.~Piro.
\newblock Diffusion of passive scalars in fluid flows: Maps in three
  dimensions.
\newblock In R.~Jullien, L.~Peliti, R.~Rammal, and N.~Boccara, editors, {\em
  Universalities in Condensed Matter}, pages 236--241. Springer, Berlin, 1988.

\bibitem[FKP88b]{Feingold88b}
M.~Feingold, L.P. Kadanoff, and O.~Piro.
\newblock Passive scalars, three-dimensional volume-preserving maps, and chaos.
\newblock {\em J. Stat. Phys.}, 50(3/4):529--565, 1988.

\bibitem[FM13]{Fox13a}
Adam~M. Fox and James~D. Meiss.
\newblock Greene's residue criterion for the breakup of invariant tori of
  volume-preserving maps.
\newblock {\em Physica D}, 243(1):45--63, 2013.
\newblock http://www.sciencedirect.com/science/article/pii/S016727891200245X.

\bibitem[FM14]{Fox14a}
Adam~M. Fox and James~D. Meiss.
\newblock Efficient computation of invariant tori in volume-preserving maps.
\newblock {\em In Review}, 2014.

\bibitem[Gor12]{Gorodetski12}
A.~Gorodetski.
\newblock On stochastic sea of the standard map.
\newblock {\em Communications in Mathematical Physics}, 309:155--192, 2012.

\bibitem[HT10]{Hogg10}
R.V. Hogg and E.A. Tannis.
\newblock {\em Probability and Statistical Inference, Eighth Edition}.
\newblock Pearson, Upper Saddle River, NJ, 2010.

\bibitem[Kar83]{Karney83}
C.F.F. Karney.
\newblock Long time correlations in the stochastic regime.
\newblock {\em Physica D}, 8:360--380, 1983.

\bibitem[Mei92]{Meiss92}
J.D. Meiss.
\newblock Symplectic maps, variational principles, and transport.
\newblock {\em Rev. of Mod. Phys.}, 64(3):795--848, 1992.
\newblock http://dx.doi.org/10.1103/RevModPhys.64.795.

\bibitem[Mei12]{Meiss12a}
J.D. Meiss.
\newblock The destruction of tori in volume-preserving maps.
\newblock {\em Communications in Nonlinear Science and Numerical Simulation},
  17:2108--2121, 2012.
\newblock http://dx.doi.org/10.1016/j.cnsns.2011.04.014.

\bibitem[MO86]{Meiss86}
J.D. Meiss and Edward Ott.
\newblock Markov tree model of transport in area-preserving maps.
\newblock {\em Physica D}, 20:387--402, 1986.

\bibitem[NM65]{Nelder65}
J.A. Nelder and R.~Mead.
\newblock A simplex method function minimization.
\newblock {\em Computer Journal}, 7:308--313, 1965.

\bibitem[RdSCV04]{Roberto04}
M.~Roberto, E.C. da~Silva, I.L. Caldas, and R.L. Viana.
\newblock Transport barrier created by dimerized islands.
\newblock {\em Physica A}, 342:363--369, 2004.

\bibitem[ST97]{Simo97}
C.~Sim\'o and D.~Treschev.
\newblock Evolution of the last invariant curve in a family of area preserving
  maps.
\newblock {\em Preprint}, 1997.

\bibitem[VM12]{Vayda12}
Umesh Vaidya and Igor Mezi{\'c}.
\newblock Existence of invariant tori in three dimensional maps with
  degeneracy.
\newblock {\em Phys. D}, 241(13):1136--1145, 2012.

\bibitem[Xia92]{Xia92}
Z.~Xia.
\newblock Existence of invariant tori in volume-preserving diffeomorphisms.
\newblock {\em Erg. Th. Dyn. Sys.}, 12(3):621--631, 1992.
\newblock http://dx.doi.org/10.1017/S0143385700006969.

\bibitem[Yoc92]{Yoccoz92}
Jean-Christophe Yoccoz.
\newblock Travaux de {H}erman sur les tores invariants.
\newblock {\em Ast\'erisque}, (206):Exp.\ No.\ 754, 4, 311--344, 1992.
\newblock S{\'e}minaire Bourbaki, Vol. 1991/92.

\end{thebibliography}

\end{document}